\documentclass[journal=jpc,manuscript=article]{achemso}
\setkeys{acs}{articletitle = true}
\usepackage[version=3]{mhchem} 
\usepackage[utf8]{inputenc}
\usepackage[T1]{fontenc}
\usepackage{babel}
\usepackage[dvipsnames]{xcolor}
\usepackage{graphicx}
\usepackage{amsmath}
\usepackage{array,multirow}
\usepackage{tabularx}
\usepackage{float}
\usepackage{wrapfig}
\usepackage{enumitem}
\makeatletter
\def\hlinewd#1{%
\noalign{\ifnum0=`}\fi\hrule \@height #1 %
\futurelet\reserved@a\@xhline}

\SectionNumbersOn
\author{M. C. B. Barbosa}
\affiliation{IFIMUP, Institute of Physics for Advanced Materials, Nanotechnology and Photonics, Department of Physics and Astronomy, Faculty of Sciences, University of Porto, Rua do Campo Alegre, 687, 4169-007
Porto, Portugal}
\author{E. Lora da Silva}
\affiliation{IFIMUP, Institute of Physics for Advanced Materials, Nanotechnology and Photonics, Department of Physics and Astronomy, Faculty of Sciences, University of Porto, Rua do Campo Alegre, 687, 4169-007
Porto, Portugal}
\alsoaffiliation{High Performance Computing Chair, University of Évora, Largo dos Colegiais 2, 7004-516 Évora}
\email{estelina.silva@fc.up.pt}
\author{P. Neenu Lekshmi}
\affiliation{IFIMUP, Institute of Physics for Advanced Materials, Nanotechnology and Photonics, Department of Physics and Astronomy, Faculty of Sciences, University of Porto, Rua do Campo Alegre, 687, 4169-007
Porto, Portugal}

\author{M. L. Marcondes}
\author{L. V. C. Assali}
\author{H. M. Petrilli}
\affiliation{Universidade de S\~{a}o Paulo, Instituto de F\'{i}sica, Rua do Matao 1371, 05508-090, S\~{a}o Paulo, SP, Brazil}

\author{A. M. L. Lopes}
\author{J. P. Ara\'{u}jo}
\affiliation{IFIMUP, Institute of Physics for Advanced Materials, Nanotechnology and Photonics, Department of
Physics and Astronomy, Faculty of Sciences, University of Porto, Rua do Campo Alegre, 687, 4169-007
Porto, Portugal}

\title{Pressure-Induced Phase Transformations of Quasi-2D Sr$_3$Hf$_2$O$_7$}

\abbreviations{Pressure,DFT,Lattice Dynamics,QHA}

\begin{document}

%
%
%
%
%

\begin{abstract}
We present an \textit{ab-initio} study of the quasi-2D layered perovskite Sr$_3$Hf$_2$O$_7$ com\-pound, performed within the framework of the Density Functional Theory and lattice dynamics analysis. At high temperatures, this compound takes a \textit{I4/mmm} centrosym\-met\-ric  structure (S.G. n. 139); as the temperature is lowered, the symmetry is broken into other intermediate polymorphs before reaching the ground state structure, which is the \textit{Cmc2$_1$} ferroelectric phase (S.G. n. 36). One of these intermediate polymorphs is the \textit{Ccce} structural phase (S.G. n. 68). Additionally, we have probed the \textit{C2/c} system (S.G n. 15), which was obtained by following the atomic displacements corresponding to the eigenvectors of the imaginary frequency mode localized at the $\mathbf{\Gamma}$-point of the \textit{Ccce} phase.
By observing the enthalpies at low pressures, we found that the \textit{Cmc2$_1$} phase is thermodynamically the most stable. Our 
results show that the \textit{I4/mmm} and \textit{C2/c} phases never stabilize in  
the 0-20 GPa range of pressure values. On the other hand, the \textit{Ccce} phase becomes energetically more stable at around 17 GPa, surpassing the \textit{Cmc2$_1$} structure. By considering the effect of entropy and the constant-volume free energies, we observe that the \textit{Cmc2$_1$} polymorph is
 energetically the most stable phase at low temperature; however, at 350 K the \textit{Ccce} system becomes the most stable. By probing the volume-dependent free energies at 19 GPa, we see that \textit{Ccce} is always the most stable phase between the two structures and also throughout the studied temperature range. When analyzing the phonon dispersion frequencies, we conclude that the \textit{Ccce} system becomes dynamically stable only around 19-20 GPa, and that the \textit{Cmc2$_1$} phase, is metastable up to 30 GPa. 
\end{abstract}

\section{Introduction}
Rational design and the discovery of new materials are essential towards the advancement of future technological innovations. Perovskite structures (with general chemical formula ABX$_3$, where A and B are typically large and small cations, respectively, and X is an anion) are one of the most versatile classes of materials, with a variety of fascinating characteristics such as magnetism, superconductivity, thermoelectricity, electrocatalysis, ferroelectricity, and optical properties. \cite{NatMat.6.296.2007,NatMat.7.425.2008} While the greater part of the known perovskite oxides are centro\-symme\-tric, a small fraction of these crystallize as non-centrosymmetric (polar perovskites).\cite{Campbell:wf5017} The polar phase enables the emergence of ferroelectricity, piezoelectricity and nonlinear optical effects which are crucial to modern electronic devices, such as sensors, actuators, tunnel junctions for non-volatile memory, transducers for medical imaging, and solar cells.\cite{NatMat.6.296.2007,NatMat.7.425.2008,doi:10.1021/jacs.5b05406,Chen2005ReviewCO,https://doi.org/10.1002/cssc.201900671} 

Among perovskites, naturally layered perovskite oxides, such as quasi-2D Ruddlesden-Popper (RP) and Dion-Jacobson (DJ) structures, have particular interest because of their recently discovered mechanism for ferroelectricity, termed as "hybrid improper ferroelectrici\-ty" (HIF). \cite{HIF,InorgChem.53.3769.2014} Structures with HIF exhibit two non-polar structural distortion modes, which lead to a third polar distortion mode, and are known as the trilinear coupling, resulting in a macroscopic spontaneous polarization.\cite{HIF,BENEDEK201211} For perovskite structures which undergo distortions towards the \textit{Pnma} space group (i.e. double-perovskite structures), the HIF phenomenon leads to an antiferro-distortive displacement of the $A$-site cations and the adjacent layer displacement in opposite directions.\cite{D0TC03161E,D1TC00989C} The displacements of the adjacent layers cancel each other for ABO$_3$ single perovskites, leading to non-polar phases. However, naturally layered quasi-2D perovskites, with disconnected octahedral layers and having two inequivalent crys\-tal\-logra\-phic A-sites, exhibit macroscopic layer polarization, \cite{doi.org/10.1002/adfm.201300210,C5DT00010F} which are oriented oppositely with slightly different intensities.
In the RP structured naturally layered oxides, with the (AO)(ABO$_3$){$_n$} stoichiometry form, the $n$-layers of the ABO$_3$ perovskite slabs are stacked  between the AO rock-salt layers, with a relative shift of the neighboring perovskite slabs by a ($\frac{1}{2}$,$\frac{1}{2}$) translation. 
Increasing the value of $n$ from 1 to $\infty$ it is possible to tune the structural dimensionality of these perovskites from quasi-2D to 3D. At high temperatures, each of the $n$ layers consists of vertex sharing oxygen octahedra with a B-ion sitting at its center. \cite{HIF,BENEDEK201211} As temperature decreases, these systems can undergo structural phase transitions due to reorientation of oxygen octahedra, leading to HIF. The HIF is also predicted to occur in oxide superlattices. \cite{Nature.452.732.2008} In addition to HIF, the octahedral reorientations in layered perovskites are of continuing interest because these eﬀects also trigger many interesting electronic properties, such as high-T$_c$ superconductivity, colossal magnetoresistance, metal-insulator transitions, and magnetic ordering.\cite{Harris}

Recent theoretical works from Liu \textit{et al.} \cite{HIF_SHO} and da Silva \textit{et al.}\cite{groupt}  confirmed that Sr$_3$Hf$_2$O$_7$ (SHO), a $n$=2 quasi-2D RP layered perovskite system, with a ferroelectric ground state \textit{Cmc2$_1$} (S.G. n. 36), is similar to the previously reported HIF systems, such as Ca$_2$Mn$_3$O$_7$ and Ca$_2$Ti$_3$O$_7$, in Ref. \citenum{Harris}. SHO is also characterized by a high-temperature \textit{I4/mmm} (S.G. n. 139) centrosymmetric paraelectric structure. Previous studies showed that by applying group theoretical analysis it was possible to probe potential intermediate polymorphs that form as the temperature is lowered, enabling a transition pathway as \textit{I4/mmm} $\rightarrow$ \textit{Cmcm} (S.G. n. 63) $\rightarrow$ \textit{Cmca} (S.G. n. 64) $\rightarrow$ \textit{Cmc2$_1$} (S. G. n. 36).\cite{HIF,groupt} The \textit{Ccce} (S.G. n. 68) structural phase is a possible polymorph, however it is not evidenced as being an intermediate structure, occurring for these structural RP symmetries, according to group theoretical analysis - a condensation of the X$_1^{-}$ zone boundary mode would have to occur upon symmetry breaking of the centrosymmetric system. It was also found by computing the phonon dispersion curves of the \textit{Ccce} phase,\cite{groupt} that the system is dynamically unstable at 0 K and 0 GPa, evidencing negative phonon modes localized at two of the high-symmetry points of the Brillouin-zone (BZ): $\mathbf{\Gamma}$- and \textbf{Y}-points. Therefore the \textit{Ccce} polymorph does not have the potential to crystallize at room conditions. We must note that such a phase has been experimentally evidenced as an intermediate phase for increasing temperatures in systems with similar stoichiometry, i.e. Ca$_3$(Mn/Ti)$_2$O$_7$, \cite{CMO, Pomiro} thus making it plausible for this polymorph to form through a first-order transition for the SHO system as well.  

Only recently have high-pressure (HP) studies been devoted to understand the properties of the perovskite systems and the respective transition pathways. Pressure is an important ther\-mo\-dynamic variable, for it enables a precise control over the interatomic distances and hence the atomic interactions, in turn driving phase-transitions. Almost all materials will undergo several phase-transitions under compression, thus generating new polymorphs with promising properties, different to those properties of materials under room conditions. It has been found that factors which are known to control the ABO$_3$ perovskites pressure response, such as the A- and B-site cation formal charges, the tolerance factor, and the B-site chemical environments, also affect the pressure response of the layered perovskites.\cite{PhysRevB.104.144105} Therefore, the interest in applying pressure to probe the structural, electronic and vibrational properties to such structures.

A recent experimental study on Sr$_3$Sn$_2$O$_7$ compound evidences a sequence of pressure-induced phase transitions as \textit{Cmc2$_1$} $\rightarrow$ \textit{Pbcn} (S.G. n. 60) $\rightarrow$ \textit{Ccce} $\rightarrow$ \textit{I4/mmm}, at room temperature.\cite{PhysRevB.104.064106} Such transition pathway would match the sequence of temperature-dependent structural transitions observed for this compound between 77 and 1000 K.
For the Ca$_3$M$_2$O$_7$-based oxides (M = Ti, Mn),\cite{Zhang.55.113001.2021} ferroelectric switching is inhibited by the irregular domains induced by the intermediate \textit{Ccce} phase, and the switching may be realized when the intermediate phase is suppressed by applying chemical pressure. Moreover, it has been evidenced that the O octahedra tilt and rotation modes of Ca$_3$Ti$_2$O$_7$ undergo softening when hydrostatic pressure or heating is considered.\cite{PhysRevB.97.094104,PhysRevB.97.041112} This behavior under temperature and pressure reveals the softness of the antiphase tilt, which indicates the importance of the partially occupied $d$-orbitals of the transition metal ions, since these determine the stability of the oxygen octahedra distortion.\cite{PhysRevB.97.094104,PhysRevB.97.041112} Moreover, through Raman measurements, it was observed that under isotropic pressure, the polyhedra tilts of the Ca$_3$Mn$_2$O$_7$ compound can be suppressed within the low pressure regime (1.4-2.3  GPa).\cite{PhysRevB.99.224105} On the other hand, through DFT calculations (0 K) performed on the Ca$_3$Ti$_2$O$_7$ compound, it was recently found by computing the enthalpies for several pressure values, up to 20 GPa, that in fact the \textit{Ccce} phase can become energetically more favorable than \textit{Cmc2$_1$}, with the transition occurring slightly below 15 GPa.\cite{PhysRevB.104.144105}

The goal of this work is to inspect, through first-principles calculations, the evolution of the structural properties, the energetic and dynamic stabilities of the \textit{Cmc2$_1$} and \textit{Ccce} structural phases of the SHO compound for increasing values of applied hydrostatic pressure. Further, enthalpy results on the \textit{I4/mmm} and \textit{C2/c} structural phases have also been computed.

To probe the energetic (thermodynamic) stability, we evaluate the enthalpy difference curves of the \textit{Ccce}, \textit{C2/c}, and \textit{I4/mmm} structural phases, relative to the ground state \textit{Cmc2$_1$} phase at 0 GPa, for hydrostatic pressures up to 20 GPa. We also analyze the free energies, where the zero-point energy is considered, in order to infer the energetic stability trend between the \textit{Ccce} and \textit{Cmc2$_1$} structures. It is worth mentioning that we have considered not only the harmonic approximation (HA), but also the quasi-harmonic approximation (QHA) at constant pressure of 19 GPa. Upon these results, the dynamic stability was probed by analyzing the phonon dispersion curves of the two energetically lowest structural phases - \textit{Ccce} and \textit{Cmc2$_1$} - at relevant pressure values.

\section{Theoretical Methodology}

Density Functional Theory (DFT) \cite{hohenberg-pr-136-1964,PhysRev.140.A1133.1965} calculations were performed through the use of the Quantum Espresso (QE) code. \cite{Giannozzi-JPhysCondensMatt-21-395502-2009,Giannozzi-JPhysCondensMatt-29-495901-2017,Giannozzi-JChemPhys-152-154105-2020} For the exchange-correlation (xc) functional, the semi-local generalized-gradient approximation with the Perdew-Burke-Ernzerhof revised for solids (PBEsol) \cite{perdew-prl-100-2008,perdew-prl-102-2009} was used.

Firstly, variable-cell relaxation calculations were performed for each structural phase (\textit{Cmc2$_1$}, \textit{Ccce}, \textit{C2/c},  and \textit{I4/mmm}) of the SHO compound and for each applied hydrostatic pressure value, ranging from 0 GPa up to 20 GPa. In these structural relaxations, both the atomic positions and the unit-cell parameters were allowed to change, by constraining the system to the targeted pressure value. The Projector Augmented Wave (PAW) \cite{PhysRevB.50.17953.1994} datasets were used to treat semi-core electronic states, with the Sr$\,[4s^2\,4p^6\,5s^2]$, Hf$\,[5s^2\,5p^6\,5d^2\,6s^2]$,  and O$\,[2s^2\,2p^4]$ electrons being included in the valence shell. Additionally, the variable cell-shape relaxation, considering the damped Beeman ionic dynamics and the Wentzcovitch extended Lagrangian for the cell dynamics,  \cite{Wentzcovitch-PhysRevB-44-2358-1991} was  performed with a plane-wave kinetic-cutoff of 70 Ry. The electronic Brillouin-Zone (BZ) was sampled with a $\mathbf{\Gamma}$-centered Monkhorst-Pack mesh \cite{monkhorst-prb-13-1976} and defined with 6 $\times$ 6 $\times$ 12 subdivisions for the \textit{Ccce} and \textit{C2/c} phases; 12 $\times$ 12 $\times$ 6 subdivisions for the ground-state \textit{Cmc2$_1$} system, and 14 $\times$ 14 $\times$ 14 subdivisions for the \textit{I4/mmm} phase.

The enthalpy curves, $H$, were then calculated by interpolating the $V-p$ and $E-V$ curves with a 3$^{rd}$ order Birch-Murnagham equation \cite{birch-pr-71-1947,murnaghan-pnas-30-1944} to obtain the relation $H = E + pV$, where $E$ is the total electronic energy of the system, $p$ is the hydrostatic pressure to which the system is subjected to,  and $V$ is the volume per formula unit. The calculation of the relative enthalpy curves, with respect to the lowest enthalpy at 0 GPa of the \textit{Cmc2$_1$} phase, allows the analysis of the evolution of the thermodynamic stability of the several structural phases, relative to one another, with increasing values of applied pressure (up to 20 GPa).

For the (harmonic) lattice-dynamics calculations, the supercell finite-displacement meth\-od was considered, with the Phonopy software package, \cite{phonopy} where QE is used as the second-order force-constant calculator. The supercells employed to compute the phonon dispersion spectra were a 2 $\times$ 2 $\times$ 2 expansion of the primitive-cell. Phonon calculations were performed for the \textit{Ccce} structure at several pressure values above which the system is energetically more favorable than the \textit{Cmc2$_1$} phase. Respective calculations were considered in order to probe the dynamical stabilities of the \textit{Ccce} phase. Phonon frequencies calculations were also considered for the \textit{Cmc2$_1$} system at pressure values between 20 and 30 GPa.

In the harmonic model, the equilibrium distance among atoms is not temperature depen\-dent. The anharmonic effects needed to account for thermal expansion can be intro\-duced by the QHA, in which the thermal expansion of the crystal lattice is obtained from the volume dependence of the phonon frequencies. \cite{PhysRevB.81.174301.2010,PhysRevB.91.144107} To perform a QHA calculation, the phonon frequencies
are computed for a range of expansions and compressions about the 0 K equilibrium volume, and the constant-volume free energy, for each configuration, is then evaluated as a function of temperature. From this approach, the equilibrium volume, bulk modulus, and Gibbs free energy can be obtained at several temperature values by fitting the free energy as a function of volume to the Vinet equation of state (EoS).\cite{PhysRevB.35.1945,PhysRevB.81.174301.2010,phonopy} 
We have performed QHA calculations for the two structural phases of interest: \textit{Cmc2$_1$} and \textit{Ccce}. The EoS fit was computed for a constant pressure of 19 GPa and for a temperature range up to 500 K (more details on the EoS fitting is discussed in Supp. Information).

\section{Results and Discussion}

\subsection{Sr$_3$Hf$_2$O$_7$ Polymorphs}

Since the environment around the Sr atoms is inequivalent, the general formula of the SHO system can be better written as (AHfO$_3$)$_2$A$'$O, where ${\rm A}'={\rm Sr}_1$ and ${\rm A}={\rm Sr}_2$. Just as other RP structures, the system results from the inter-growth of rock-salt (R) and perovskite (P) blocks. The P blocks are composed of two layers of HfO$_6$ octahedra along the $a$-axis, sharing the O corner ions. The A$'$-site cations occupy a nine-fold coordination site (R block), while the A-site cations have a coordination number of twelve (cuboctahedral in P block). 

The high-symmetry structure is centrosymmetric and belongs to the \textit{I4/mmm} space group (S. G. n. 139).\cite{groupt} By decreasing the temperature, lower symmetry structural phases may be generated by inducing tiltings and/or rotations of the O octahedra cages. In Ref. \citenum{groupt} it was theoretically shown that second-order phase transitions may occur, towards lower structural phases, to the \textit{Cmcm} or \textit{Cmca} space groups, from which the phonon instability modes of these two phases may direct the final transition to \textit{Cmc2$_1$} space group. Interestingly enough, it was experimentally observed that other related RP systems, such as Ca$_3$Mn$_2$O$_7$ and Ca$_3$Ti$_2$O$_7$ compounds, a structural transition path from low-temperature \textit{Cmc2$_1$}, intermediate-temperature \textit{Ccce} (S.G. n. 68) and high-temperature \textit{I4/mmm} occurs.\cite{CMO} Since the \textit{Ccce} phase is not related by group-symmetry analysis to the \textit{I4/mmm} $\rightarrow$ \textit{Cmc2$_1$} transition, since there is no single mode connecting the phases, we infer that this structural phase may occur as a first-order phase transition. Such a transition can occur by applying an external perturbation, such as pressure, in order to induce the condensation of the X$_1^{-}$ zone boundary mode from the centrosymmetric system.\cite{groupt} 

\begin{figure}[!h]
    \centering
    \includegraphics[width=14cm]{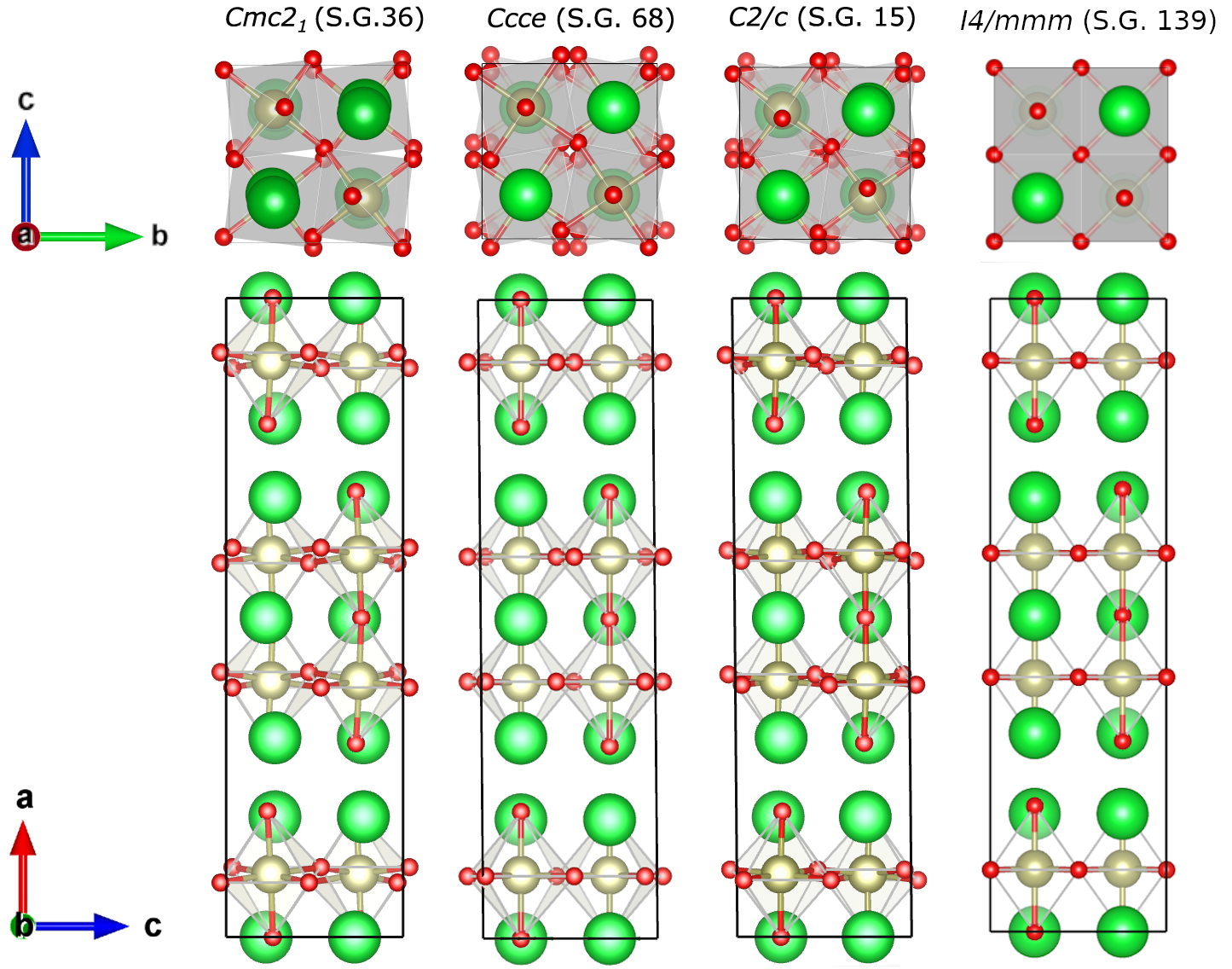}
    \caption{Representation of the \textit{Cmc2$_1$}, \textit{Ccce}, \textit{C2/c}, and \textit{I4/mmm} structural phases of Sr$_3$Hf$_2$O$_7$. The red, green, and gold spheres represent, respectively, the oxygen, strontium, and hafnium ions. 
    The VESTA visualization software was used to plot the figures. \cite{JApplCrystallogr.44.1272.2011}}
    \label{structures}
\end{figure}

We show in Fig. \ref{structures} the four structural phases we have studied in this investigation. These are:

\begin{enumerate}
\item The ground-state structural phase at room conditions, evidencing a polar symmetry, \textit{Cmc2$_1$} (S.G. n. 36). This system phase is related to the centrosymmetric \textit{I4/mmm} structure through group-subgroup relations, by the condensation of the X$_2^{+}$ and the X$_3^{-}$ modes,  which respectively lower the high-symmetry to the \textit{Cmcm} and \textit{Cmca} phases.\cite{groupt}
\item The \textit{Ccce} structural phase (S.G. n. 68), which has experimentally been observed in other similar RP compounds;\cite{CMO} however such a phase has not been evidenced in SHO by analyzing the group-subgroup relations from the \textit{I4/mmm} towards the polar \textit{Cmc2$_1$} symmetry, and as described in Ref. \citenum{groupt}. Respective polymorph is induced through the condensation of the X$_1^{-}$ mode, which does not occur spontaneously for the SHO system. Such a distortion mode would have to be induced by applying an external perturbation (i.e. hydrostatic pressure) to break the symmetry from \textit{I4/mmm} towards \textit{Ccce}. We must note that there is no group-subgroup relation (by considering single mode analysis) between the \textit{Ccce} phase and the \textit{Cmc2$_1$} phase.\cite{groupt}
\item The \textit{C2/c} system (S.G. n. 15) was obtained by mapping out the anharmonic potential energy surfaces by following the eigenvectors associated with the soft (im\-ag\-i\-nar\-y) phonon mode, localized at the $\mathbf{\Gamma}$-point, of the \textit{Ccce} structure (a more detailed description has been added in Supp. Information). From this analysis it is possible to obtain a lower energy structure, corresponding to the minima of the potential energy surface.
\item And finally, the high-symmetry \textit{I4/mmm} structural phase.
\end{enumerate}

All three structural polymorphs are related to the high-symmetry, high-temperature, \textit{I4/mmm} phase through tiltings and/or rotations of the O octahedra. When external per\-tur\-ba\-tion is applied, symmetry-breaking occurs, thus lowering the energy of the system.

\subsection{Energetic Stability}

\begin{figure}[!h]
\centering
\includegraphics[width=16cm]{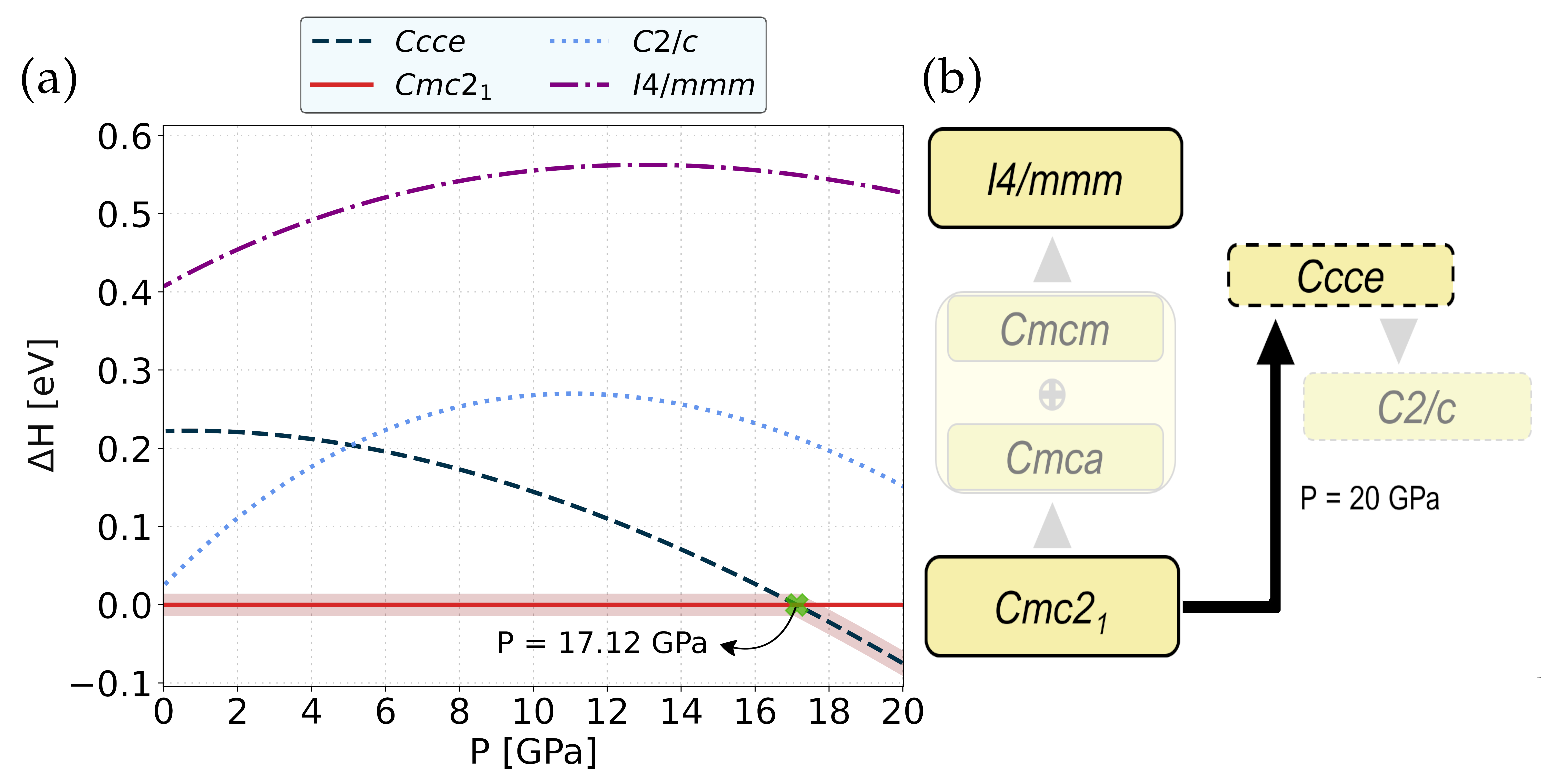}
\caption{(a) Relative enthalpy curves of the 
\textit{I4/mmm}, \textit{Ccce}, \textit{C2/c}, and \textit{Cmc2$_1$} structural phases as a function of pressure, with respect to the structure with the lowest energy value at ambient pressure (\textit{Cmc2$_1$}). The cross marker indicates the pressure value at which the \textit{Ccce} structural phase becomes energetically stable. (b) Schematics of the most plausible transition pathways that Sr$_3$Hf$_2$O$_7$ can undergo, and based also on Ref. \citenum{groupt}.}
\label{H-P}
\end{figure}

To probe whether the structural phases of Fig. \ref{structures} can become energetically competitive as a function of hydrostatic pressure, the enthalpy energies, for pressures up to 20 GPa, have been calculated. These results are presented in Fig. \ref{H-P}, where the relative enthalpy energies, $\Delta H$, with respect to that of the structure with lowest enthalpy at 0 GPa (the \textit{Cmc2$_1$} phase) are defined.

By analyzing Fig. \ref{H-P}, we observe that at 0 GPa, the orthorhombic \textit{Cmc2$_1$} phase is energetically the most stable, as already discussed in literature.\cite{groupt} The \textit{C2/c} is slightly higher in energy than \textit{Cmc2$_1$} ($\Delta$H=$\sim$0.02 eV) at 0 GPa, however starts increasing the relative enthalpy as a function of pressure up until $\sim$10 GPa. After 12 GPa the energetics of this system starts decreasing, however never becomes energetically stable throughout the studied pressure range. It is noteworthy of mentioning, that the energetic tendency to decrease for increasing pressures, evidences the possibility of the \textit{C2/c} system stabilizing for pressures higher than 20 GPa. At 20 GPa, we observe a enthalpy difference of the \textit{C2/c} system with respect to \textit{Cmc2$_1$} of $\Delta$H=0.15 eV.

At room conditions, the \textit{Ccce} phase is, in terms of enthalpy, the second highest system with respect to the ground-state structure. However, for increasing pressure we observe a considerable decrease in energy. At $\sim 5$ GPa this system will be competing energetically with the \textit{C2/c} structure, thus becoming more stable than the latter, and above 17.12 GPa the \textit{Ccce} phase becomes energetically the most favorable among the four studied systems. 

The high-symmetry \textit{I4/mmm} system shows quite elevated energy at 0 GPa, which is expected since it is a high-temperature phase. At around 10 GPa, we observe that the \textit{I4/mmm} structural phase starts decreasing the enthalpy and, at 20 GPa the energetic difference with respect to the \textit{Cmc2$_1$} phase is around 0.52 eV. In comparison to the most stable phase at 20 GPa, the \textit{Ccce} structure, the energy difference between the two phases is of $\sim 0.60$ eV. Once again, we envisage the possibility of the high-symmetry \textit{I4/mmm} system to stabilize at much higher pressures (> 30 GPa), than those considered in the present work.

The enthalpy calculations were performed at 0 K, without taking into consideration the contributions to the free energy from the vibrations of the solids. In order to probe whether these effects could alter the energy ordering between the \textit{Cmc2$_1$} and \textit{Ccce} phases of Sr$_3$Hf$_2$O$_7$, we have performed lattice-dynamics calculations on the equilibrium and compressed struc\-tures to evaluate the constant-volume Hemholtz (F) and constant-pressure Gibbs (G) free energies.

Within the harmonic approximation, the Helmholtz free energy as a function of tem\-per\-a\-ture can be obtained from the phonon frequencies and the lattice energy of the equilibrium structure. Temperature effects can be included through the Helmholtz free energy, which by introducing a transformation from the constant volume function, we obtain the thermal properties at constant pressure, and thus the Gibbs free energy. \cite{PhysRevB.91.144107,PhysRevLett.117.075502}  
By increasing tem\-per\-a\-ture, the volume dependence of the phonon free energy changes, which in turn results in different equilibrium volumes for different temperatures. This is regarded as thermal expansion in the QHA.

\begin{figure}[!t]
\centering
\includegraphics[width=11cm]{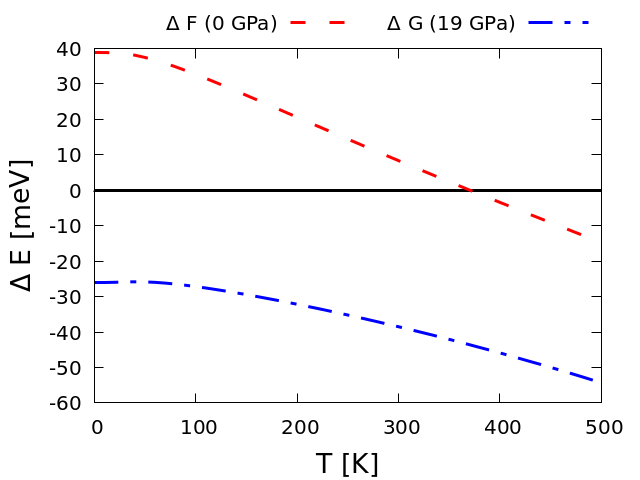}
\caption{The harmonic (constant-volume/Helmoltz; red) and quasi-harmonic (constant-pressure at 19 GPa/Gibbs; blue) free energies of the \textit{Ccce} phase relative to \textit{Cmc2$_1$} (black) as a function of temperature.}
\label{Free_energy}
\end{figure}

The temperature dependence of the relative Helmholtz, $\Delta$F (0 GPa), and Gibbs, $\Delta$G (19 GPa), free energies of the \textit{Cmc2$_1$} and \textit{Ccce} phases are shown in Fig. \ref{Free_energy}. 

We observe from the Helmoltz free energy (Fig. \ref{Free_energy}; red dashed line) that the \textit{Ccce} structure becomes energetically more stable than the \textit{Cmc2$_1$} system (Fig. \ref{Free_energy}; black solid line), surpassing the latter at 370 K. This observation leads to the conclusion that the \textit{Ccce} may also be induced through temperature effects, with the transition temperature being quite close to room temperature. At 0 K and 0 GPa, the free-energy difference between both phases is 38 meV, which is significantly lower than the relative enthalpy obtained without considering the entropy effects ($\sim$220 meV).

Since the \textit{Ccce} structure becomes dynamically stable around 19-20 GPa (more detailed information in Subsec. \ref{sec:dynamical_stability}), and that the \textit{Cmc2$_1$} structure is still stable up until 30 GPa (Subsec. \ref{sec:dynamical_stability}), we have calculated the Gibbs free energy at 19 GPa (Fig. \ref{Free_energy}; blue dashed-dotted line) to probe the evolution of the free energies of the two phases of interest. We observe that \textit{Ccce} is thermodynamically the most stable phase throughout the studied temperature range (up to 500 K). At 0 K the free energy of the \textit{Ccce} phase differs from the \textit{Cmc2$_1$} system with a relative value of $\Delta$E$\sim $25 meV; as temperature increases this energy difference also grows further apart ($\Delta$E=54 meV at 500 K). When comparing $\Delta$G with $\Delta$H (Fig. \ref{H-P}) for the two systems at 0 K and 19 GPa, we observe that the enthalpy difference is $\sim$50 meV, which is in much closer agreement to when comparing the relative enthalpy with $\Delta$F (for 0 GPa and 0 K). 

We must note that the (quasi-)harmonic approximation will not account for the influence of the soft modes of the \textit{Ccce} system on the free energy, a deficiency which is amplified by the sensitivity of the volume to the free energy. Therefore the transition temperature might be slightly underestimated than expected.\cite{PhysRevB.91.144107} 
Since at 19 GPa, neither the two analyzed phases evidence imaginary modes, we conclude that the QHA free energies would provide a much reasonable description of the free energy behavior between the two systems. Overall, and by comparing the free energies of Fig. \ref{Free_energy} to what was obtained for the relative enthalpy plots of Fig. \ref{H-P}, we observe comparable behaviors between the energetics of both polymorphs: at 0 GPa and low temperatures the \textit{Cmc2$_1$} system is more stable than \textit{Ccce}; whereas with pressure (19-20 GPa) the \textit{Ccce} is more stable than \textit{Cmc2$_1$}, even for finite temperatures.

\subsection{Dynamical Stability} \label{sec:dynamical_stability}

Energetic stability is a necessary, but not a sufficient, condition for a structural phase to be synthetically accessible. Another condition that should be analyzed is the dynamical stability of the system, which requires the study of the phonon spectra. If imaginary frequencies emerge (usually represented by negative frequencies along the phonon dispersion curves), such a feature would indicate that the system is at a transient state, 
undergoing a phase transition and thus cannot be kinetically stable at the given temperature and/or pressure conditions.\cite{ProcCambridgePhilosSoc36.160.1940,Dove.IntLattDyn,Dove.StrutDyn,AmMin.82.213.1997, BullMaterSci.1.129.1979, PhysRevLett.111.025503.2013}

\begin{figure}[!t]
    \centering
    \includegraphics[width=16cm]{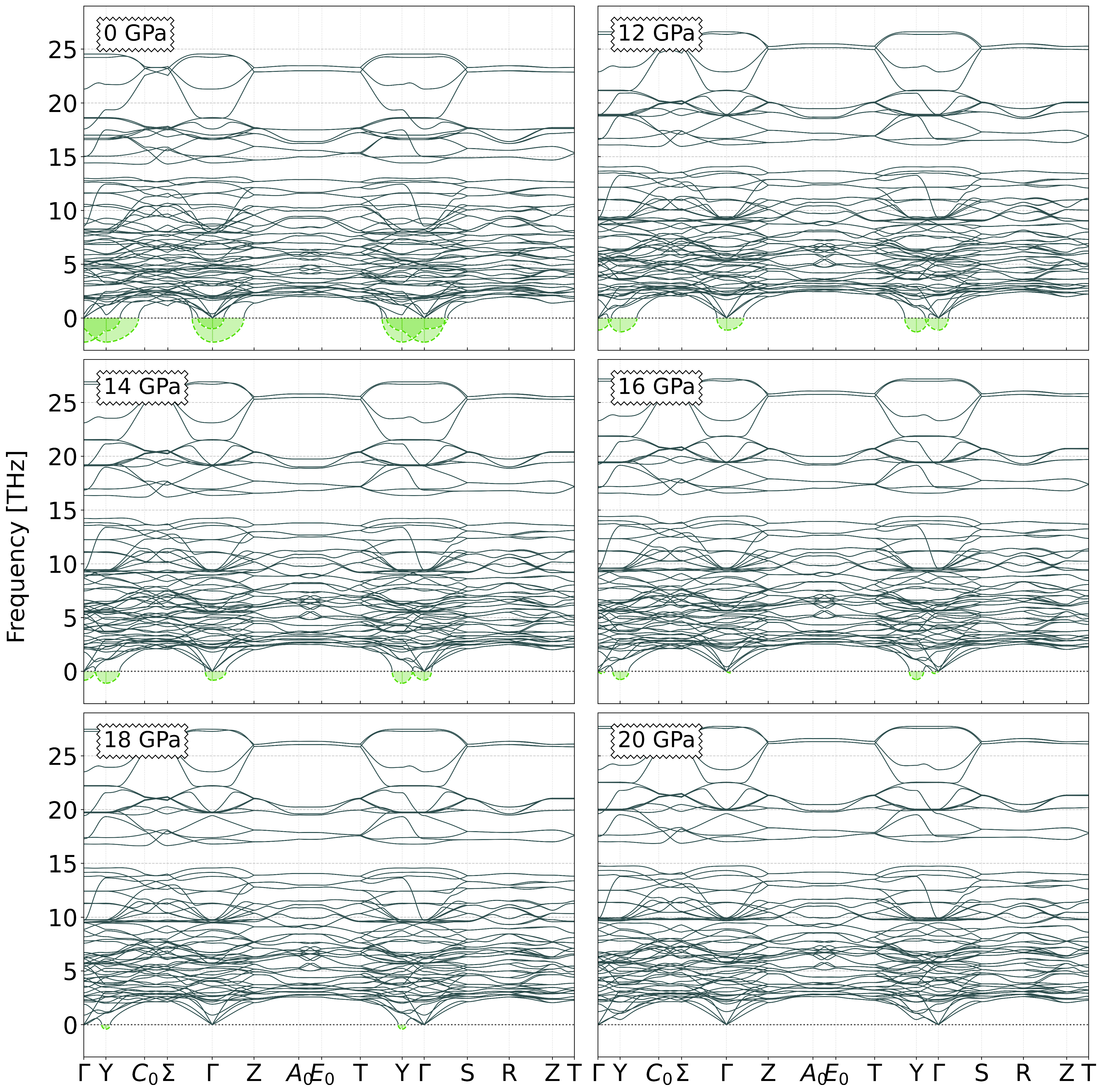}
    \caption{Phonon dispersion curves for the \textit{Ccce} structural phase of the SHO system for several pressure values: 0 GPa (taken from Ref. \citenum{groupt}), 12 GPa, 14 GPa, 16 GPa, 18 GPa, and 20 GPa.}
    \label{Phonon_Ccce}
\end{figure}

Thus being, we have proceeded to calculating the phonon  structure for the \textit{Ccce} phase at several pressure values. These values were considered at: room pressure (and taken from Ref. \citenum{groupt} for comparison); and at values above which the energetically stability of the \textit{Ccce} phase occurs - 12, 14, 16, 18, and 20 GPa, as shown in Fig. \ref{Phonon_Ccce}.

As illustrated in Fig. \ref{Phonon_Ccce}, the phonon dispersion curves of the \textit{Ccce} structural phase at 0 GPa (retrieved from Ref. \citenum{groupt}) show negative phonon modes localized at two of the high-symmetry points of the BZ: $\mathbf{\Gamma}$- and \textbf{Y}-points. We can therefore infer that such a structural phase is not kinetically stable at ambient pressure. However, we observe that as the applied pressure increases, the frequencies of the negative modes shift towards positive values. The $\mathbf{\Gamma}$-point soft-mode hardens more rapidly than the \textbf{Y}-point: at 18 GPa only a mild instability is observed at the latter point, since at $\mathbf{\Gamma}$ the soft optical mode already evidences positive frequencies. However, it is only at 20 GPa that we arrive at a structure that is both thermodynamically and dynamically stable, with the \textbf{Y} phonon mode evidencing positive values as well.
\begin{figure}[t!]
    \centering
    \includegraphics[width=16cm]{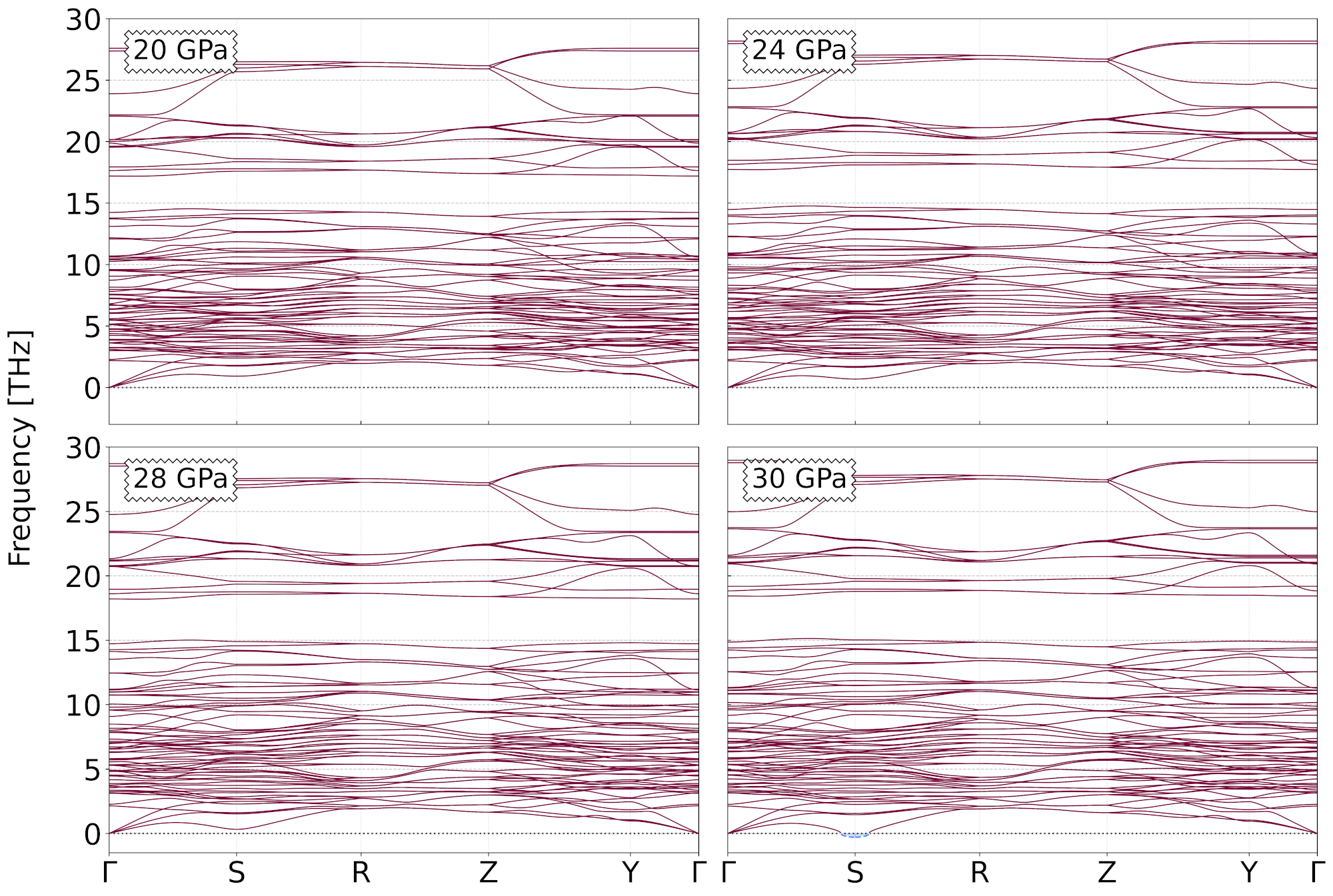}
    \caption{Phonon dispersion curves for the \textit{Cmc2$_1$} structural phase of the SHO system for high pressures: 20 GPa, 24 GPa, 28 GPa, and 30 GPa.}
    \label{Phonon_Cmc21}
\end{figure}

We also show, in Fig. \ref{Phonon_Cmc21}, the phonon dispersion curves  of the \textit{Cmc2$_1$} system, at several pressure values, and we observe that the structure is dynamically stable still at 20 GPa, thus coexisting with \textit{Ccce} phase, as a metastable compound. It is noteworthy of mentioning that around the high-symmetry \textbf{S}-point, the lowest frequency branch starts to decrease towards a soft mode, which may be an indication that for higher pressure values a structural transition will occur due to the instability at this \textbf{q}-point. Indeed, by increasing the pressure to 30 GPa 
it is observed that this soft-mode becomes unstable, thus rendering this structure kinetically unviable above this pressure value.


\section{Conclusions}

In the present work we have performed \textit{ab-initio} calculations to study the thermodynamical stability of the room temperature \textit{Cmc2$_1$} phase and the high-temperature \textit{I4/mmm} structure; as well as different structural phases of Sr$_3$Hf$_2$O$_7$, which are not related to the aristotype polymorph through group-subgroup relations, i.e., are a result of first order transitions. The latter were the \textit{Ccce} and \textit{C2/c} structural phases.

It was observed that, at room conditions, the \textit{Cmc2$_1$} phase is the most energetically stable, as expected; the \textit{C2/c} phase being 0.05 eV higher in energy than the ground-state; followed by the \textit{Ccce} system with an energy difference of $\sim 0.18$ eV with respect to \textit{Cmc2$_1$}; and finally the \textit{I4/mmm} polymorph with the highest energy difference of $\sim$0.41 eV in relation to the ground state phase. We observe that the enthalpy of  \textit{Ccce}  decreases as pressure increases, and at $\sim 17$ GPa, it surpasses that of the \textit{Cmc2$_1$} structure, becoming energetically the most stable system of all four studied compounds. 

To inspect in detail the two lowest energetic compounds, namely the \textit{Ccce} and \textit{Cmc2$_1$} phases, we have calculated the free energies (where entropy is accounted for) and the phonon dispersion curves to analyze their  respective dynamical stability. From the constant-volume free energies, we observe that the \textit{Cmc2$_1$} system is energetically the most stable compound below 370 K. Above this temperature, the \textit{Ccce} polymorph has its free energy lowered, thus becoming more stable than \textit{Cmc2$_1$} phase. By probing the volume-temperature dependent free energies at 19 GPa, we observe that the \textit{Ccce} is always the most stable phase when compared to \textit{Cmc2$_1$}, throughout the studied temperature range, up to 500 K. When observing the phonon dispersion spectra,  we conclude that the \textit{Ccce} structure only becomes dynamically stable at 20 GPa, which is at a higher pressure range than when the energetic transition occurs from \textit{Cmc2$_1$}. The \textit{Cmc2$_1$} phase, on the other hand, is dynamically stable up to 30 GPa, when an imaginary mode at the high-symmetry \textbf{S}-point emerges. These results suggest that both phases may coexist energetically at pressure values from $\sim 19$ GPa up to $\sim 30$ GPa; at this pressure interval the polar \textit{Cmc2$_1$} structure is a metastable system.

From the present calculations we conclude that the quasi-2D layered Ruddlesden-Popper perovskite Sr$_3$Hf$_2$O$_7$ compound has a stable phase at room conditions which is a ferroelectric ground state system with \textit{Cmc2$_1$} symmetry. By applying hydrostatic pressure this phase may undergo a transition to the \textit{Ccce} structure, for it becomes thermodynamically (at $\sim 17$ GPa) and dynamically stable (at 20 GPa). 

\begin{acknowledgement}

This research was supported by the Network of Extreme Conditions Laboratories (NECL), POCI-01-0145-FEDER-029454
(FLIP), POCI-01-0145-FEDER-032527 (vdW4device), \-UIDB/\-04968/\-2020 (IFIMUP - Financiamento Plurianual da Unidade de I\&D), funded by the Por\-tu\-guese Foundation for Science and Technology (FCT) and co-financed by NORTE 2020, through the programme Portugal 2020 and FEDER. Authors also thank the financial support of the FAPESP projects 2018/07760-4 and 2019/07661-9, and CNPq projects 314884/2021-1 and 311373/2018-6. E.L.d.S. acknowledges the CEEC individual fellowship 5th edition with Reference 2022.00082.CEECIND financed by FCT; and the High Performance Computing Chair - a R\&D infrastructure (based at the University of Évora; PI: M. Avillez), endorsed by Hewlett Packard Enterprise, and involving a consortium of higher education institutions, research centers, enterprises, and public/private organizations. 
The authors acknowledge PRACE for awarding access to the Fenix Infrastructure resources, which are partially funded from the European Union’s Horizon 2020 research and innovation programme through the ICEI project under the grant agreement No. 800858 (technical support provided by CINECA). Additionally, the authors acknowledge the national computer resources from FCT CPCA/\-A2/\-7087/\-2020 at MACC - Bob, and CPCA/A2/7263/2020 and CPCA/A1/431509/2021 (2ª Edição) at Univ. Évora - Oblivion, respectively. 

\end{acknowledgement}

%
%

\section{Supplementary Material}

\subsection{Evolution of the lattice parameters with applied pressure}

\begin{figure}[!h]
    \centering
    \includegraphics[width=14cm]{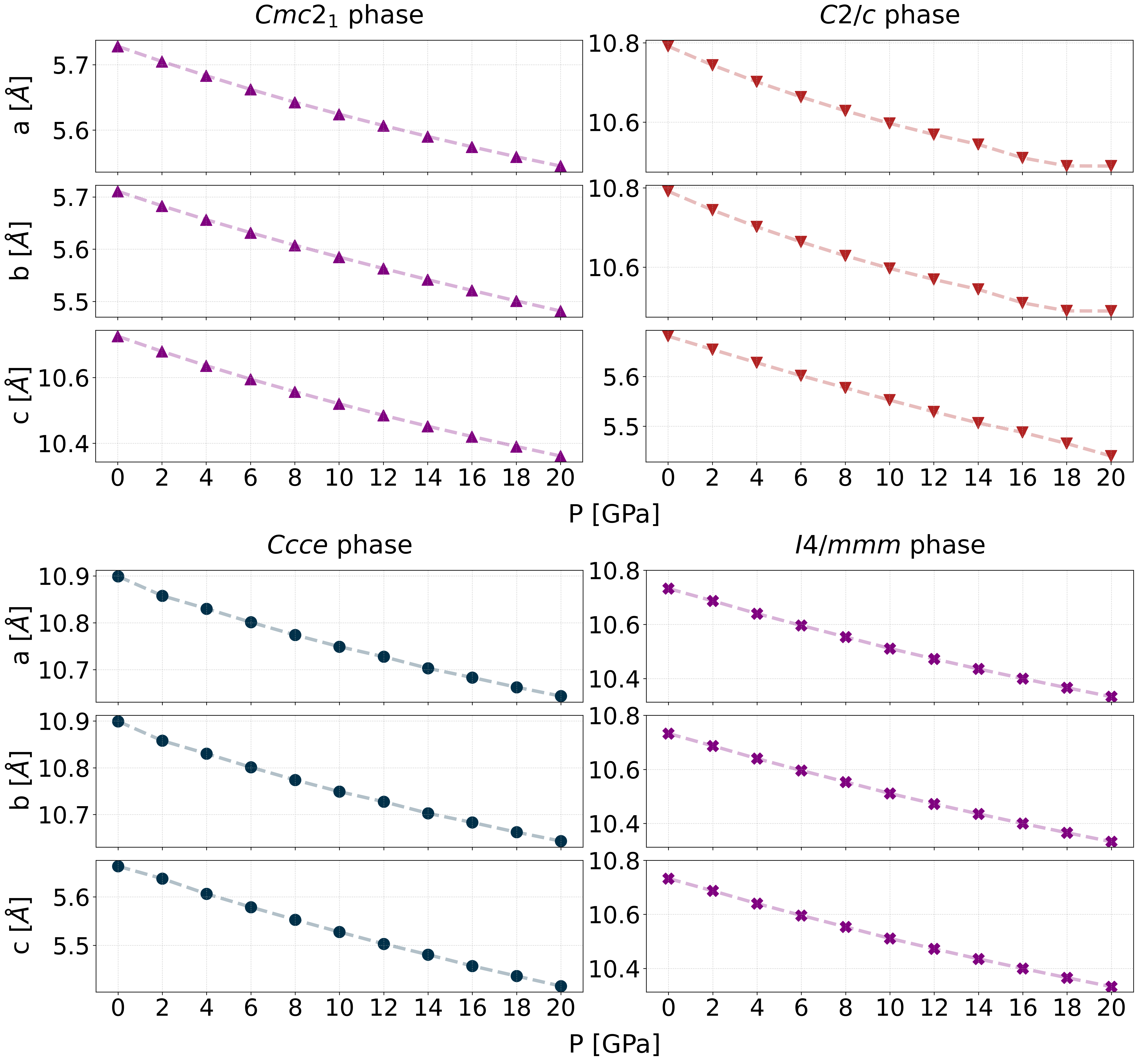}
    \caption{Evolution of the calculated lattice parameters ($a$, $b$, and $c$) for the four of the considered structures of the SHO system with increasing hydrostatic pressure. }
    \label{lattice_const}
\end{figure}

In Fig. \ref{lattice_const} we show the lattice parameters of the four SHO studied systems as a function of pressure in the 0-20 GPa range, namely the \textit{Cmc2$_1$}, the \textit{C2/c}, the \textit{Ccce}, and the \textit{I4/mmm} phases.

\clearpage
\subsection{Potential-Energy Surface}
\label{sec:pot}

\begin{figure}[!h]
    \centering
    \includegraphics[width=14cm]{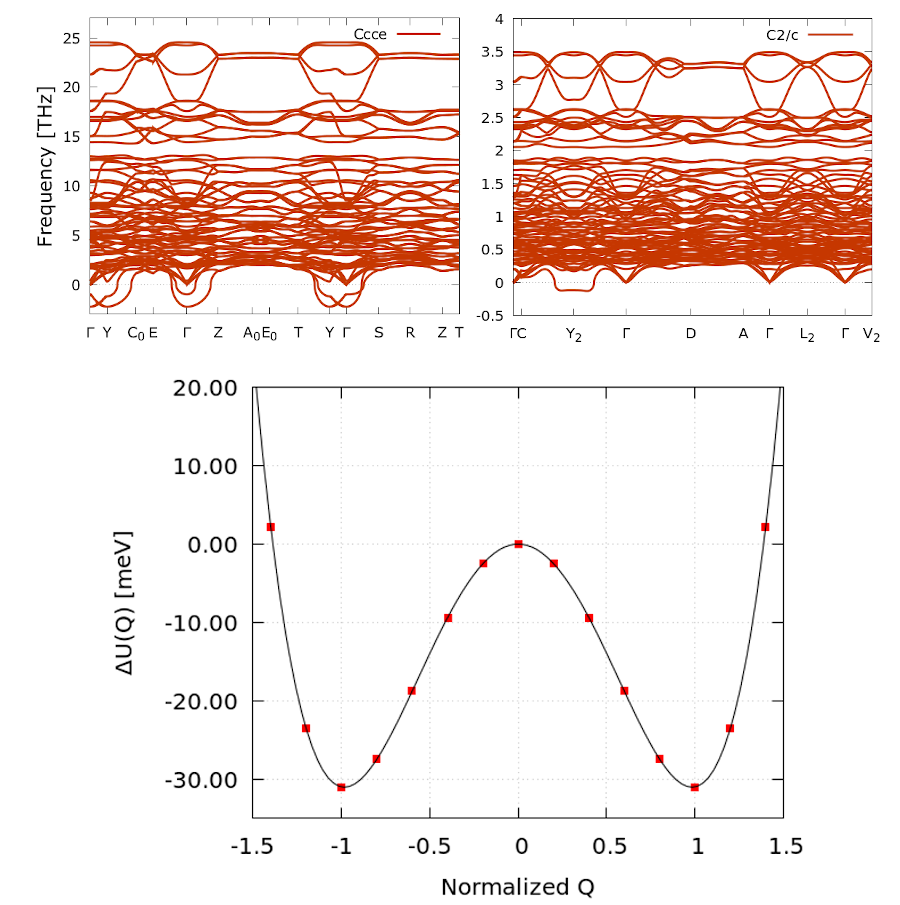}
    \caption{Double-well potential-energy surface (bottom plot) for the phonon instabilities associated with the high-symmetry $\mathbf{\Gamma}$-point negative mode of the \textit{Ccce} phase [top, left figure (results of the phonon dispersion are from Ref. \citenum{groupt})]. The minima of the potential-energy surface correspond to the energetically favorable structural phase at 0 GPa, the \textit{C/2c} polymorph, where the respective phonon modes are also shown (top, right plot). The normal-mode coordinates \textit{Q} have been normalized so that the minima are located at \textit{Q}=$\pm$1 and the energy differences are those calculated in a unit-cell.}
    \label{pot_surf}
\end{figure}

By analyzing the \textit{Ccce} phonon dispersion curves of SHO at 0 GPa (Fig. \ref{pot_surf}, top, left) we observe two unstable modes localized at the high-symmetry $\mathbf{\Gamma}$-point. Considering the frequency mode with higher degree of instability (larger negative frequency), we have mapped out the anharmonic potential energy surfaces by following the eigenvectors associated with this instability/displacement. From this analysis, it is possible to obtain a lower energy structure, corresponding to the minima of the potential energy surface. This displacement involves the O-cage distortions and the Sr displacements at the unit-cell, and minimizing the the energy down to -30.97 meV (\textit{Q}=1), by lowering the symmetry to \textit{C2/c}, when compared to the \textit{Ccce} phase at \textit{Q}=0 (Fig. \ref{pot_surf}).

For these calculations we used the open-source MODEMAP package.\cite{PhysRevLett.117.075502,modemap} A sequence of displaced structures is generated by displacing along the phonon eigenvectors over a range of amplitudes of the normal-mode coordinate \textit{Q}. The total energies of the “frozen phonon” structures are then evaluated from single-point DFT calculations, after which the E(\textit{Q}) curves will be fitted to a polynomial function, with the number of terms depending on the form of the potential well. 

By computing the phonon dispersion structure of the energetically favorable system, the \textit{C2/c}, we observe that it is dynamically unstable at room conditions (Fig. \ref{pot_surf}, top, right). A negative mode appears at the \textbf{Y}$_2$ \textbf{q}-point, which is related to the \textbf{Y} mode from the \textit{Ccce} structure.
 
 \clearpage

\subsection{Electronic Band Structure of the \textit{Ccce} and \textit{Cmc2$_1$} Structural Phases}

\begin{figure}[!h]
    \centering
    \includegraphics[width=16cm]{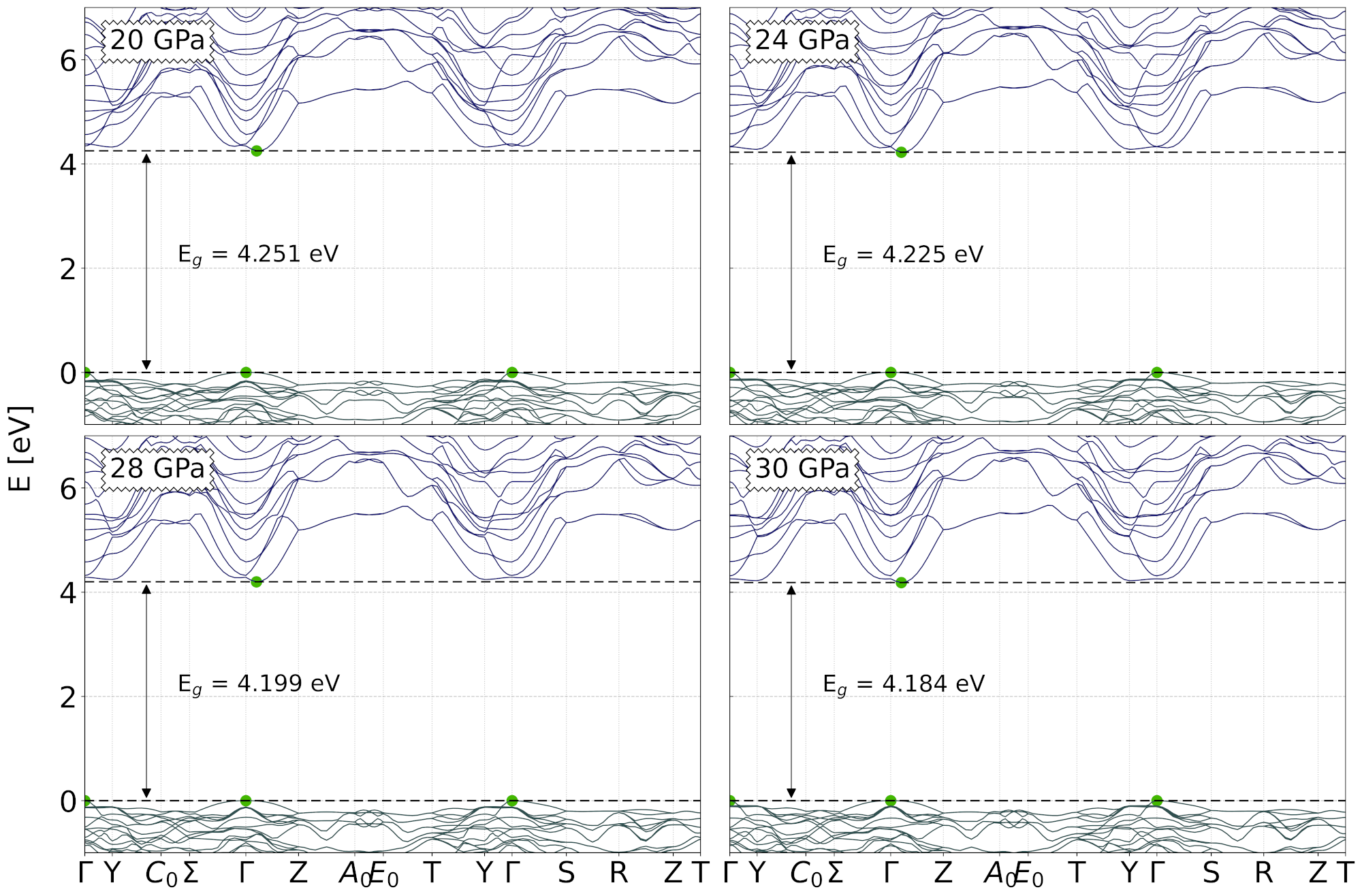}
    \caption{Electronic band structures of the \textit{Ccce} structural phase of the SHO system at several applied pressure values: 20 GPa, 24 GPa, 28 GPa, and 30 GPa.}
    \label{bands_Ccce}
\end{figure}

The electronic band dispersions at pressure values above which the dynamical stability has been confirmed for the \textit{Ccce} structure is presented in Fig. \ref{bands_Ccce}. These values  were chosen to be 20 GPa, 24 GPa, 28 GPa, and 30 GPa. We may infer that as pressure increases, a subtle decrease of the band-gap occurs, from 4.25 eV (at 20 GPa) towards 4.18 eV (at 30 GPa). All structures display an indirect band-gap, where the valence band maximum (VBM) is located at the $\mathbf{\Gamma}$-point, and the conduction band minimum (CBM) is slightly shifted away from the $\mathbf{\Gamma}$-point and along the $\mathbf{\Gamma}$-\textbf{Z} segment. It should be noted, however, that for the \textit{Ccce} structural phase (at all considered pressure values), the difference between direct and indirect band-gaps is small (less than $\sim 0.1$ eV) and within the DFT error margin. Thus being, we may infer that most optical absorption will be direct.

The same procedure has been adopted for the \textit{Cmc2$_1$} phase, and we observe an opposite trend with respect to the \textit{Ccce} system: a tendency to increase the band-gap energy for increasing pressures. The \textit{Cmc2$_1$} structure evidences also an indirect band-gap, except at 12 GPa. The location of the VBM varies for different pressure values; however it is noteworthy of mentioning that the extremal points throughout the BZ are very close in energy, and therefore are within the DFT error margin. While for 0 GPa, the VBM is located at the \textbf{G$_0$}-\textbf{T} high-symmetry segment, at 12 GPa the VBM is along $\mathbf{\Gamma}$-\textbf{Z}. For larger pressures, the VBM is locked along the \textbf{Z}-\textbf{B$_0$} high-symmetry segment. The CBM seems to be more stable under different pressure values, remaining located at the $\mathbf{\Gamma}$-point. The gap increases from 4.27 eV (at 0 GPa) to 4.52 eV (at 28 GPa).

\begin{figure}[!t]
    \centering
    \includegraphics[width=16cm]{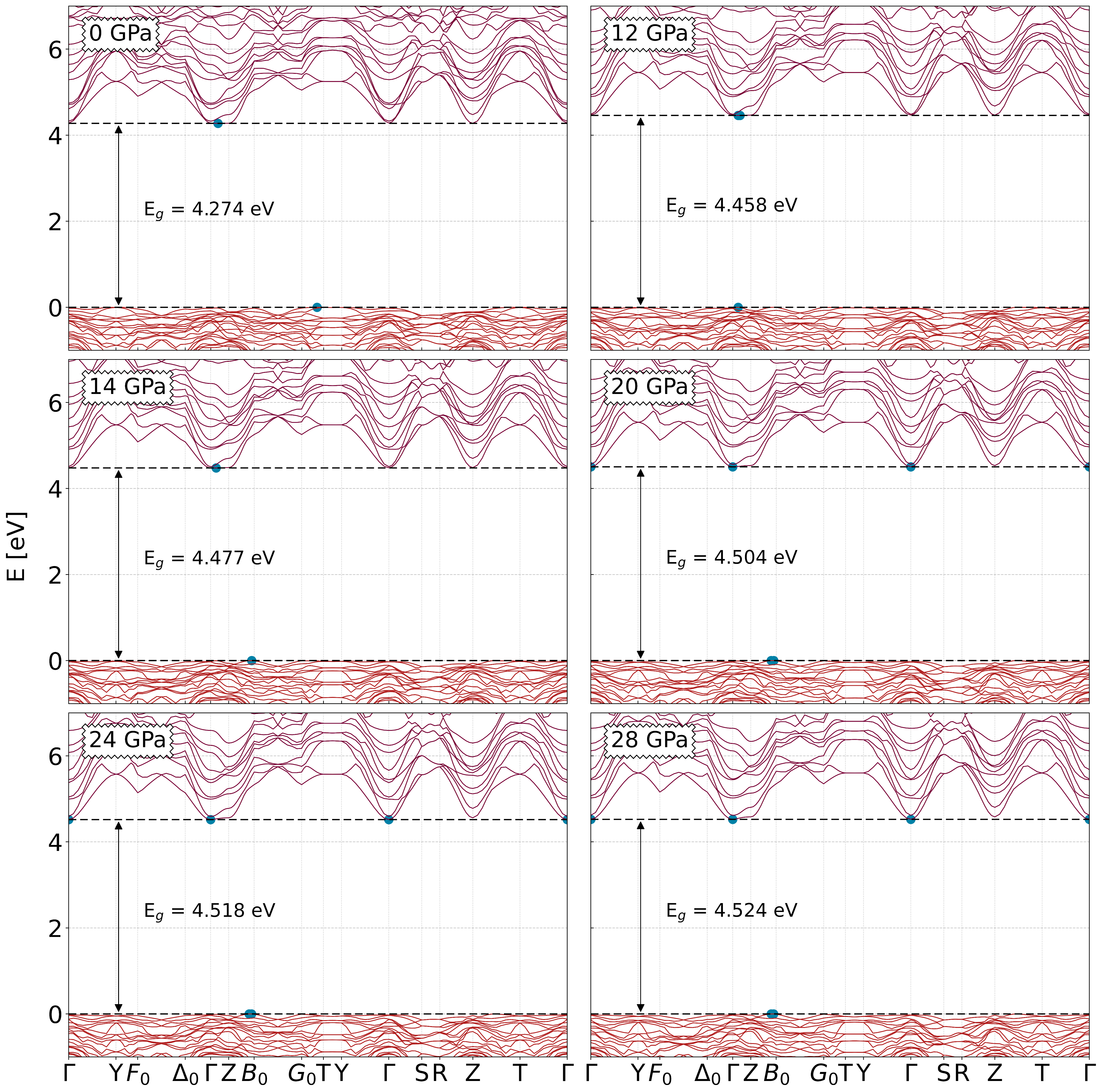}
    \caption{Electronic band structures of the \textit{Cmc2$_1$} structural phase of the SHO system at different applied pressure values: 0 GPa (taken from Ref. \cite{groupt}, 12 GPa, 14 GPa, 20 GPa, 24 GPa, and 28 GPa.}
    \label{bands_Cmc21}
\end{figure}

We have found in literature, and from experimental data of the parent perovskite systems, that at room temperature these may yield band-gap values around $\sim$ $6.1$ eV;\cite{noh_choi_lee_cho_jeon_lee_2013, sousa_rossel_marchiori_siegwart_caimi_locquet_webb_germann_fompeyrine_babich_etal_2007, rossel_sousa_marchiori_fompeyrine_webb_caimi_mereu_ispas_locquet_siegwart_etal_2007} however, theoretical DFT calculations yield lower values [as expected due to the well-known problem of the asymptotic behavior when using (semi-)local functionals]. From photoluminescence measurements, it was observed two visible emission lines close to 420 and 500 nm in AZrO$_3$ and AHfO$_3$ (A = Ca, Sr, and Ba), with respective spectral weights being sensitive to the local distortions.\cite{noh_choi_lee_cho_jeon_lee_2013} Moreover, calculations for the high-symmetry and high-temperature, cubic \textit{Pm$\overline{3}$m} structural phase of the parent systems returned band-gaps ranging from $3.7$ eV to $5.79$ eV. \citep{wang_zhong_wang_zhang_2001, hou_2008, fabricius_peltzer_y_blanca_rodriguez_ayala_de_la_presa_lopez_garcia_1997, feng_hu_cui_bai_li_2009, osti_1208247} These values were obtained through DFT calculations, by employing LDA and GGA for the xc functional. For room temperature and the  \textit{Pnma} structural phase, the calculations resulted in a $4.18$ eV band-gap width. \citep{osti_1206659}

Tuning and/or engineering of the band-gaps towards lower values for double-layered RP systems could in fact be possible, either through cation mutations and/or replacing O by other chalcogen elements (i.e. S, Se), since these may alter the Fermi energy accordingly, by changing the position of the CBM and/or VBM, respectively. This study is however out of the scope of the present work, and therefore we will not enter into further discussion.

\subsection{Electronic Band Structure of \textit{Cmc2$_1$} with Spin-Orbit Coupling Effects}
\label{sec:soc}

\begin{figure}[!h]
    \centering
    \includegraphics[width=16cm]{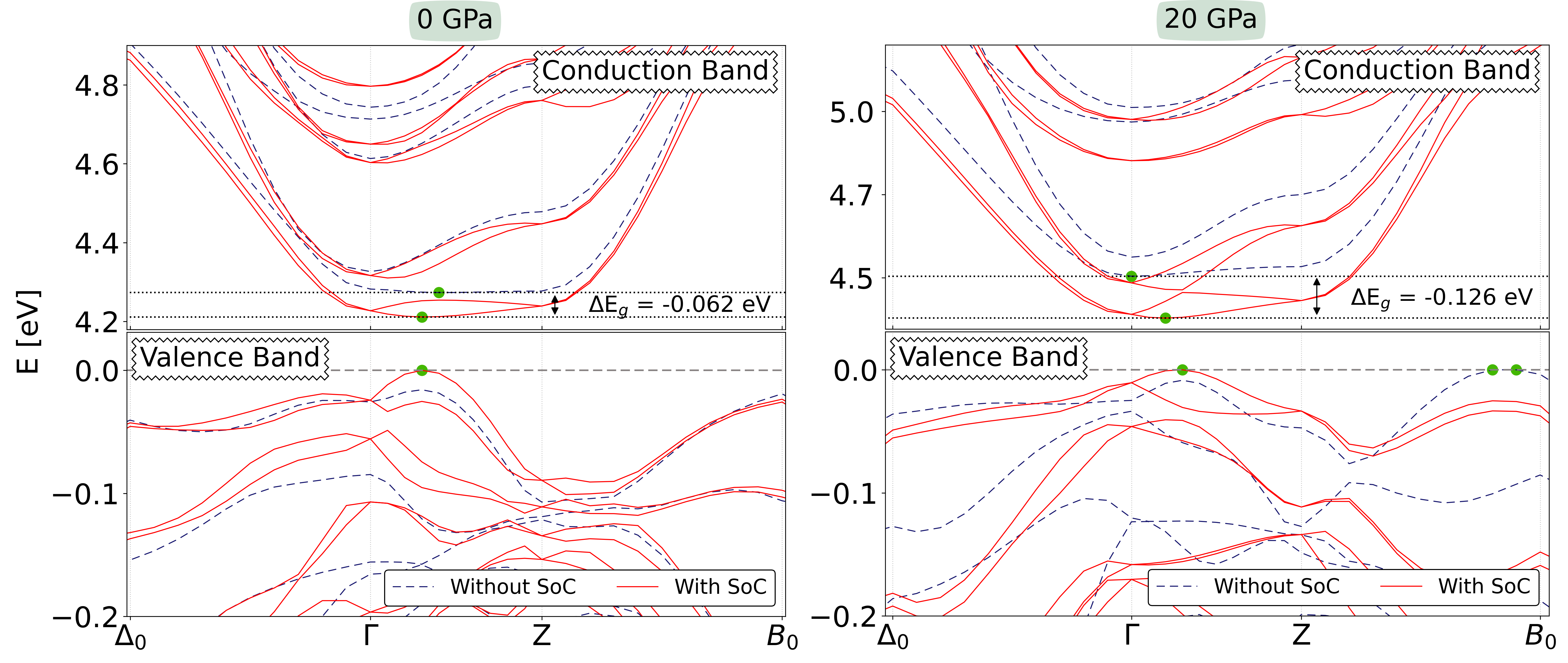}
    \caption{Fully relativistic electronic band structure (solid red line) of the conduction band minimum, compared to the system without considering the spin-orbit interaction (purple dashed line) of the polar \textit{Cmc2$_1$} structure at 0 GPa (top) and 20 GPa (bottom). }
    \label{soc_cbm}
\end{figure}

We compare in Fig. \ref{soc_cbm} the electronic band dispersion of the \textit{Cmc2$_1$} system when taking the Spin-Orbit Coupling (SoC) effects into account, and when these interactions are not considered in the calculations. We see that at 0 GPa the energy splitting of the CBM is only 0.062 eV, with an increase of the band energy splitting occurring at 20 GPa with 0.126 eV. 

On the other hand, for the VBM, we observe that the position of the band maximum changes. For room pressure (0 GPa), the VBM goes from the high-symmetry segment \textbf{G}$_0$-\textbf{T}, when we do not consider SoC, to a direct character, when SoC is included, with the VBM aligning with the CBM at the same high-symmetry point, along the segment $\mathbf{\Gamma}$-\textbf{Z}. When pressure is applied with 20 GPa, we also observe a change of position of the VBM from the \textbf{Z}-\textbf{B}$_0$ high-symmetry segment to $\mathbf{\Gamma}$-\textbf{Z}, although remaining with an indirect character, as to what occurs without relativistic effects. We must state that the changes of the VBM position when considering, and not considering, SoC, are related to the fact that the valence band extrema along the BZ are very close to the Fermi energy, having energy maxima very similar to each other, and therefore these changes are quite subtle. 

From these calculations we do not observe any major effects of the electronic properties when fully relativistic calculations are considered, and most importantly these calculations do not evidence any momentum-dependent band spin splitting for the studied polar structural phase. 

\clearpage

\subsection{Quasi-Harmonic Approximation}
\label{sec:qha}

\begin{figure}[!h]
    \centering
    \includegraphics[width=16cm]{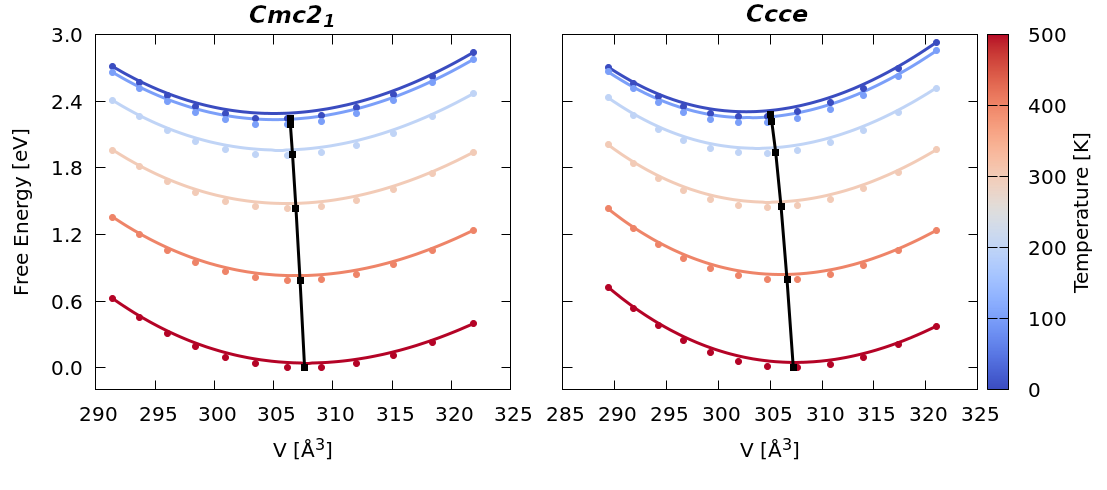}
    \caption{Helmoltz free-energy equation-of-state fits for the two Sr$_3$Hf$_2$O$_7$ polymorphs at different lattice temperatures: 
\textit{Cmc2$_1$} (left), and \textit{Ccce} (right), for a pressure value of 19 GPa and maximum temperatures up until 500 K. Note that the free energies are given as relative energies with respect to the lowest energy-volume point at 500 K.}
    \label{QHA_fits}
\end{figure}

Since at 19 GPa the \textit{Cmc2$_1$} and \textit{Ccce} polymorphs are both dynamically stable, we have probed the QHA derived-properties of the two systems at this pressure range. Moreover, at this pressure value, we obtain good fittings of the free-energy equations of state for both the studied phases, as can be observed in Fig. \ref{QHA_fits}. Since at low pressures the \textit{Ccce} phase evidences several imaginary frequencies (more detailed information in Subsec. 3.3), we found anomalies in the low pressure free-energy equations of state, which were only resolved when fixing the pressure close to 19 GPa.

\begin{figure}[!h]
    \centering
    \includegraphics[width=16cm]{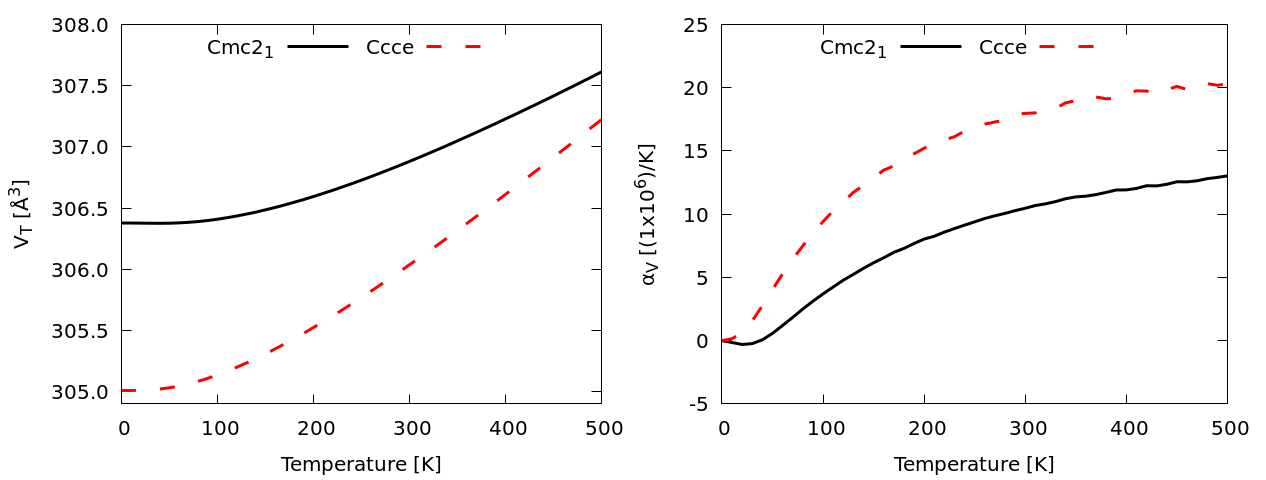}
    \caption{Volume per formula unit, $V(T)$ (left), and volumetric-expansion coefficient, $\alpha_\textrm{V}(T)$ (right), for the two Sr$_3$Hf$_2$O$_7$ polymorphs: \textit{Cmc2$_1$} (black), and \textit{Ccce} (red), for a pressure value of 19 GPa and maximum temperatures until 500 K. }
    \label{QHA_volume}
\end{figure}

The structural properties, as a function of temperature, obtained from the QHA are shown for each phase in Fig. \ref{QHA_volume}. These include the temperature-dependent volume (per formula unit), $V(T)$ and the volumetric thermal-expansion coefficient, $\alpha_V(T)$. 

The temperature-induced volume changes in SHO are dependent on the stretching of the Hf-O bonds and the tilting of the HfO$_6$ octahedra. 
As the \textit{Cmc2$_1$} and \textit{Ccce} structures have a similar HfO$_6$ octahedral framework, one may infer that the octahedral cages remain largely ideal (i.e., undistorted), and so the thermal expansion is accommodated primarily by changes in the volume of the cage.\cite{PhysRevB.91.144107} 
We observe that the volume expansion (Fig. \ref{QHA_volume}, left) of the \textit{Ccce} polymorph has a larger increase than \textit{Cmc2$_1$} when the temperature raises, which makes the former system a more compressible structure. Moreover, the thermal expansion
of the \textit{Ccce} phase (Fig. \ref{QHA_volume}, right) is found to be higher than that of the lower-symmetry phase, for increasing temperatures, however showing a similar trend at temperatures above 50 K. For the low temperature regime, the expansion coefficients of the \textit{Cmc2$_1$} system decreases slightly to negative values, indicating negative thermal expansion (NTE), i.e. a reduction in volume upon heating. The NTE can be associated to several microscopic mechanisms, however for this specific system to ferroelectricity, since this is in fact a polar phase.\cite{Miller.JMatSci.44.5411.2009}  Above 50 K, a rapid increase in the expansion coefficients are evidenced for \textit{Cmc2$_1$} until $\sim$150 GPa, followed by a slower increase up until 500 K. Such a behavior is also noticed for \textit{Ccce}, although with a much steeper increase between the interval of 50 K and 150 K.

\clearpage

\end{document}